# CTMaaS: An innovative platform for C-ITS-enabled dynamic Traffic and Fleet Management as a Service

Areti Kotsi[1], Vasileia Klimi[1], Dr.Evangelos Mitsakis[1]

[1]Centre for Research and Technology Greece (CERTH) - Hellenic Institute of Transport (HIT)
E-mail: akotsi@certh.gr, vklimi@certh.gr, emit@certh.gr

**Abstract**

Fleet management systems have been one of the most important research fields in transportation science. Nowadays the enhancement of fleet management systems with technologies such as the Cooperative Intelligent Transport System (C-ITS) that allows fleets to communicate with their environment, with other vehicles or with the road infrastructure, resulting in safer and more efficient road travel. This paper aims to present the CTMaaS platform, a tool which integrates C-ITS services and traffic management processes to manage vehicle fleets. Starting with a literature review, the paper presents various fleet management systems, that have been developed in the last years, and the most typical C-ITS services. The next chapters present the CTMaaS platform, use cases, and methodology.

## *1. Introduction*

The 21st century is characterized by the rapid growth of megacities and urbanization in Europe and worldwide. This results in the increase of urban transport demand and in the growth of the number of vehicles that load the urban and peri-urban road networks. In this context, congestion is emerging in all medium and large cities and is evolving at a rate of 3% per year (Tom Tom , 2023). At the same time, e-commerce increases the number of fleets and freight trips in urban areas. In addition, the post-Covid-19 period, that cities go through nowadays, has caused serious traffic and mobility problems that make the need for their management imperative (Andreas Nikiforiadis, 2020).

Efficient traffic management constitutes a priority concerning travel time and traffic conditions improvement in urban networks as well as for achieving sustainable mobility goals (Luke Butler, 2020). More specifically, innovative, and environmentally friendly traffic management processes are on the focus as congestion is the major environmental pollutant in cities (Ninad Lanke, 2013). New innovative technologies and management techniques have been applied in modern traffic management systems in the last decade. These include Cooperative Intelligent Transport Systems (C-ITS), sensors and systems for real-time traffic monitoring, traffic and demand forecasting based on big data analytics, and Artificial Intelligence (AI)-based traffic and network management techniques. Various commercial solutions, such as navigation systems and fleet management applications, were developed with a focus on using traffic conditions information to enable users' reactions to congestion.

Integration of the above technologies and advanced traffic management techniques are applied to achieve proactive and intelligent traffic management. An intelligent traffic management system aims to manage traffic effectively during emergencies using cutting-edge communication and processing technologies and appropriate intelligent system algorithms. When systems that manage road networks' capacity (traffic management and control systems) interact with demand management systems (fleet management systems), the level of success in decreasing congestion is expected to be high.

The objective of this paper is to present an innovative platform for dynamic traffic and fleet management developed within the context of the CTMaaS project. The paper provides a thorough review of existing fleet management systems and elaborates on the specific functionalities of the CTMaaS platform. More specifically, the use cases, the integrated C-ITS services, and the architecture of the platform are presented. Finally, conclusions are provided and aspects for future research on the domain are discussed.

*1.2 Literature review*

In the past years, various Fleet Management Systems (FMSs) were developed worldwide, aiming to provide optimized routing schedules, improve the accuracy of deliveries, and ensure safety (Ricardo Salazar-Cabera, 2020). SigFlot is an intelligent FMS enabling the fleet manager to overview the vehicle's status using dynamic information and real-time data. The system uses the Client-Server architecture and specific modules have been developed. The main module manages the consistency of the data. There are two databases; the first one stores the information needed for the identification of the fleet and the second one records the geographic information from the road network. The "Interface module" allows the manager to edit the system's database and track all the requests from the clients. The "Web module" connects the system to the internet. The "Communications and Location Module" provides a connection between the base station and the onboard units. The "SMS Service module" sends alerts in the form of free text to the clients. Another module has a map interface where the manager can enter information into the system. Finally, the "Optimizing module" distributes the assignment of tasks to be performed by the drivers, to make the system more efficient (D.B. Giralda, 2006).

An FMS developed by Ellegood et al. (William A. Ellegood, 2020) focuses on the school bus routing problem (SBRP) and covers the sub-problems of bus stop selection, bus route generation, bus route schedules, school bell time adjustment, and strategic transportation policy issues. Safety, cost-efficiency, and more effective use of the school bus fleet are some of the main targets of the system. Parameters like the capacity of the buses, the service environment (urban or rural), the maximum number of stops, and the total student walking distance are considered for calculating the best routing schedules.

The intelligent Ambulance FMS developed by Chhabria et. al (Monica Chhabria, 2016) organizes the whole process from the moment the client asks for an ambulance till the moment he/ she is transferred to the hospital. The user asks for an ambulance through the application, the system detects the exact location of the patient, and the shortest path to the hospital is calculated using real-time data from the Google Map API. The ongoing traffic situation is considered crucial for the on-time arrival of the ambulance both to the patient and to the hospital. That is why the system ensures constant communication with the traffic control room so that they can modify the traffic situation. There are four modules in the system's basic structure: Registration, Ambulance Booking, Request Processing, Placement, and Analysis. The information that the user shares for his/ her registration in the system is very important in case of an emergency. The location of the patient is given via GPS. The system guides the ambulance to the nearest hospital that can address the specific emergency. Finally, the goal of the proposed solution is to reduce the response time to every request and to increase the punctuality of the emergency response systems.

A comparative analysis of the abovementioned systems and the CTMaaS platform regarding their functionalities is presented in the table below. The CTMaaS platform functionalities mentioned in the table are thoroughly described in the next sections of the paper.

*Table 1: Comparative analysis of the CTMaaS platform and other FMSs*

| Functionalities | CTMaaS | SigFlot | School Bus Routing Problems (SBRP) FMS | Intelligent Ambulance FMS |
|---|---|---|---|---|
| C-ITS services for highways environment | X | | | |
| Prediction capabilities | X | | | |
| Priority in signaled intersections | X | | | |
| Data storage | X | | | X |
| Real-time vehicle tracking | X | | | X |
| Routing & rerouting | X | | X | X |
| ETA | X | | | X |
| Real-time Data | X | X | | |
| Road network visualization | X | X | | |
| Drivers' tasks allocation | X | X | | |
| Communication with the Traffic Control Center | X | | | |

An FMS has, in general, the objective of ensuring the safe transport of products and personnel, keeping its processes simple, designing a smart routing plan for the fleet, and adapting its actions in emergency situations (Ricardo Salazar-Cabera, 2020), (Yi-Chung Hu, 2015), (Vasileios Zeimpekis). Technological advances in location, navigation, and communication systems have made this possible targeting to enhance FMS in a dynamic manner where raw data provided by roadside units and road sensors is processed in real-time offering essential information that leads to improved customer services and reduced delays (Amelia C. Regan, 1998), (Elyes ben Hamida, 2015). This will be facilitated with technologies that connect vehicles that exchange data with other vehicles and digital infrastructure on the road. This bidirectional communication is achieved with C-ITS services which use two types of wireless communication technologies: ITS-G5 and LTE-V2X (4G/5G).

C-ITS are a group of Intelligent Transport Systems (ITS) technologies that enable the exchange of data from the road environment and can improve safety, traffic, energy efficiency, and comfort (Meng Lu, 2018), (Ling Sun, Architecture and Application Research of Cooperative Intelligent Transport Systems, 2016), (Muhammad Naeem Tahir, 2021). For example, when there is a need for changing the route of a vehicle due to traffic congestion, a Road-side Unit (RSU) utilizes CAM messages from approaching vehicles, which are used to identify the low speed of vehicles and alert the service provider, who selects then the suitable C-ITS service that applies in the situation and provides the driver with advice through an IVIM message. The driver acts accordingly, avoiding traffic congestion and/ or a potential accident (Areti Kotsi E. M., 2020). The least complex C-ITS services are called "Day 1 services" (Froetscher Monschiebl, 2018). The next phases of the C-ITS deployment include the "Day 1.5" and "Day 2" services which are related to more sophisticated and complex environments, such as the urban and peri-urban sections of the road network. These services are presented in the table below.

*Table 2:* List of Day 1 and Day 1.5 C-ITS Services

| Type | C-ITS Service |
| --- | --- |
| Day 1: Signage applications | In-Vehicle Signage (IVS) |
| | Probe Vehicle Data (PVD) |
| | Road Hazard Warning (RHW) |
| | Signal Violation |
| | Traffic signal priority for designated vehicles |
| | Green Light Optimal Speed Advisory (GLOSA) |
| Day 1: Hazardous location notifications | Emergency Brake Light |
| | Emergency Vehicle Approaching |
| | Slow or Stationary Vehicle |
| | Traffic Jam Ahead Warning |
| | Road Works Warning (RWW) |
| | Weather conditions |
| Day 1.5 | Information on fueling and charging stations |
| | Vulnerable road user protection |
| | On and off-street parking information |
| | Park and ride information |
| | Connected and cooperative information |
| | Traffic information and smart routing |

Concerning the most typical C-ITS services:

- RWW provides warnings about road works that may be either mobile or static, short, or long-term, and affect the road layout and the driving regulations. The service informs the road user when he/ she is approaching a work zone and provides simultaneous information on the changes in the road layout. The objectives are traffic flow improvement and accidents decrease.
- RHW provides information related to one or several potentially hazardous events on the road, such as stationary vehicles, weather conditions warnings, or obstacles on the road. The road user is informed about the location and the type of hazard.

- IVS service informs users about actual static or dynamic road signs.
- GLOSA provides drivers with optimal speed advice when they approach a signalized intersection. This service takes advantage of real-time traffic sensing and infrastructure information, which can then be communicated to the vehicle.
- PVD is a service in which vehicle or road user data are collected by the road operator or service provider. Connected vehicles broadcast their position, speed, and direction at any time and other vehicle information, e.g., type and length.
- Smart Routing informs road operators and road users in real-time about the traffic conditions and facilitates the selection of the most appropriate traffic management decision or route.

## 3. CTMaaS project

### 3.1 Overview

The CTMaaS project aims to develop an integrated dynamic traffic management system that utilizes innovative technologies for the coordinated and large-scale provision of C-ITS services and the implementation of dynamic traffic management and control strategies and measures. The innovation of the CTMaaS project relies on the combination of the use of C-ITS services with the optimization of traffic management and control, an innovative approach when considering the existing systems. CTMaaS provides two types of services:

- Fleet management.
- C-ITS Services for urban and highway networks.

The CTMaaS platform, which is the system developed within the context of the project, will be tested in real-life conditions in the road network of Thessaloniki. The figure below presents the locations in the city of Thessaloniki which are included in the project pilot.

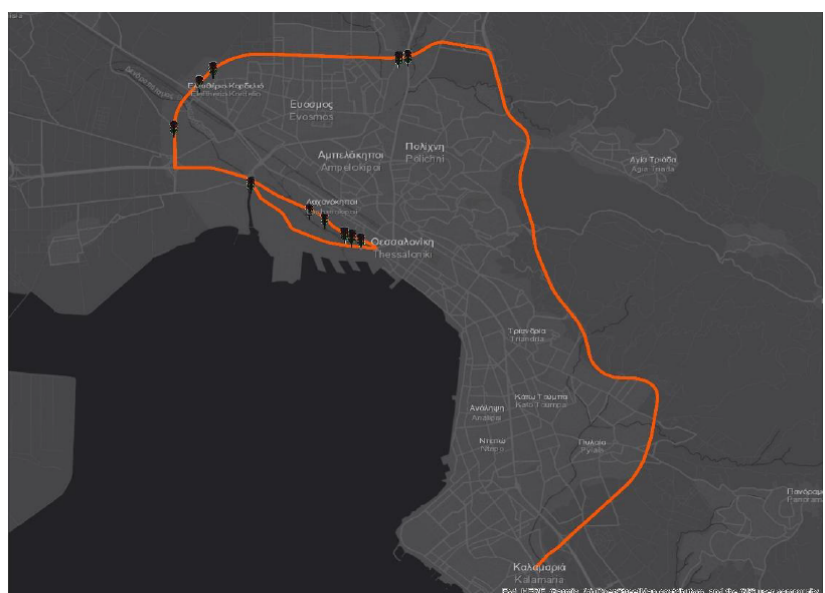

*Figure 1:* *CTMaaS pilot locations in the city of Thessaloniki*

- 5 -

*3.2 Methodology*

The goals of the CTMaaS project include the development of a system which integrates dynamic traffic management and control technologies, fleet management functionalities, and C-ITS services. The anticipated benefits are comprised of enhanced accessibility of an area, reduced environmental impacts, improved road safety, increased traffic efficiency, and overall improvement of the quality of life in the relevant area/ environment.

To achieve the goals, a Lean and Agile Development methodology has been followed for the development of the CTMaaS platform, aiming to ensure validated and verified development of the use cases within the specific implementation period. The design phase included the identification of the functional requirements of the platform through the elaboration of use cases and the definition of its architecture. The development phase, which is currently ongoing, includes the development of the individual subsystems, their integration, as well as the development of the necessary interfaces. The parameterization phase includes the integration of the C-ITS services, intelligent traffic signaling, and highway/ motorway traffic management. The following figure summarizes the methodology approach of the CTMaaS project.

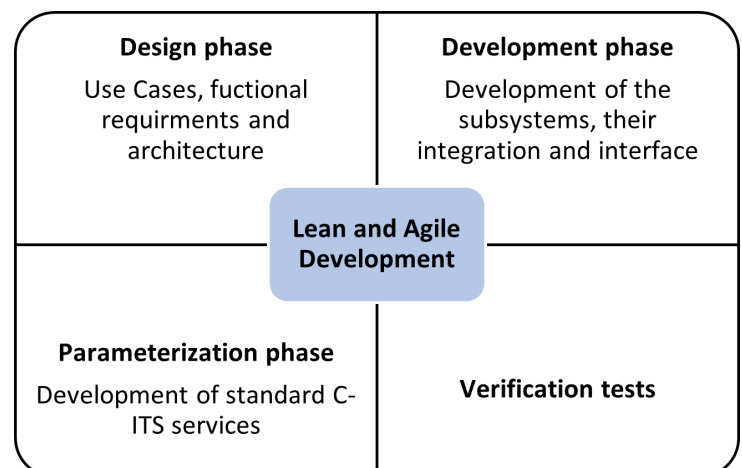

*Figure 2:* CTMaaS methodological approach

More specifically, use cases were developed to describe the actual scenarios and actions that can be performed by the platform and are directly related to the requirements of the system. The use cases are designed to express the functionalities that the platform should offer through its operation. The use cases elaborate thoroughly on all the actions that the user of the platform can perform. For each use case, there is a corresponding technical or general requirement that specifies the actions that should be performed, to achieve the desired outcome. Considering the related requirements, the architecture of the platform was designed. After these steps, verification tests will be carried out, to verify the quality and functionality of the platform.

*3.3 Use Cases*

The use cases describe the scenarios that the platform executes and the operations it performs. The following table lists the eight use cases related to the Fleet Management services of the CTMaaS platform.

*Table 3: Use Cases for Fleet Management*

| No | Use Case | Actors |
| --- | --- | --- |
| 1 | General Fleet data entry | Fleet manager |
| 2 | Entry of fleet destinations | Fleet manager |
| 3 | Dynamic vehicles fleet visualization | Fleet Manager, Driver |
| 4 | Estimated Time of Arrival (ETA) | Fleet Manager, Driver |
| 5 | Dynamic Routing | Fleet Manager, Driver |
| 5.1 | Rerouting | Fleet Manager, Driver |
| 5.2 | Change of routing with driver initiative | Driver |
| 6 | Storage of statistical data/ vehicle fleet data | Fleet Manager |
| 7 | Data exchange between the Fleet Manager and the Traffic Management Center | Fleet Manager, Traffic Manager |
| 8 | Priority to fleet vehicles at signalized intersections | Fleet Manager, Traffic Manager, Driver |

The General Fleet Data Entry use case enables the fleet manager to enter data such as the number of vehicles in the fleet, the names and phone numbers of the drivers, license plate numbers, and vehicles' color. The main objective of the Entry of fleet destinations use case is the input of the addresses or coordinates of the destinations for each vehicle and their display on a map, as well as the input of the tasks that the driver should perform in these destinations (e.g., pick up, delivery, maintenance). The Dynamic fleet vehicle visualization use case enables the fleet manager to monitor visually the fleet vehicles in real-time. The Estimated Time of arrival (ETA) use case accurately calculates the estimated time of arrival of the vehicles at every destination. The Dynamic Routing use case aims to visualize the real-time traffic conditions, provide the most efficient routing for the driver to follow, and suggest rerouting of the journey if necessary (e.g., due to traffic or unexpected road closures). The Storage of statistical data/ vehicle fleet data and Data exchange between the Fleet Manager and the Traffic Management Center use cases summarize the data that the platform will store or exchange with a Traffic Management Center. These are vehicle fleet data that will be collected from the CTMaaS mobile application, which sends CAM messages to the platform's back-office system. The vehicle fleet data includes longitude, altitude, speed, orientation, timestamp, vehicle ID, trip ID, and driver ID. The statistical data calculated from the original fleet data are trip distance, trip duration, trip maximum and minimum speed, trips per vehicle, trips per driver, vehicle working hours, and drivers' working hours. The Priority to fleet vehicles at signalized intersections use case enables the request and the provision of priority to a vehicle approaching an intersection.

### *3.4 C-ITS Services*

*3.4.1 Highways networks*

The C-ITS services related to the highway's network include RWW, RHW, and IVS. The next table presents the services and the specific use cases.

*Table 4:* *C-ITS services and use cases for highways environment.*

| C-ITS Service | Use Case |
|---|---|
| RWW | Lane closure and other restrictions |
|  | Mobile road works |
|  | Planned road works |
|  | Long-term road works (e.g., for a month, a year, etc.) |
|  | Unplanned road works |
| RHW | Weather conditions warning |
|  | Obstacle on the road |
|  | Stationary vehicle |
|  | Embedded VMS (Variable Message Sign) Free Text |
| IVS | Traffic congestion |
|  | Embeddes VMS (Variable Message Sign) Free Text |
|  | Traffic Congestion |

### 3.5 CTMaaS Architecture

The design approach for the development of the C-TMaaS platform is that of Microservices. Every application using the Microservices technique is a compilation of smaller services operating individually but always communicating in the system (Astudillo, 2018). The association of these subsystems is the final product. Every different service of architecture performs a different function (Khaled Alanezi, 2022). Individual parts can be developed simultaneously by different software development teams speeding up the process. The system developed with the Microservices approach has the advantages of easy scalability, easy maintenance, and independence from specific technologies and frameworks.

Every microservice has a specific role in the platform. Regarding the C-TMaaS platform, there are six microservices for every group of functionalities that can be performed:
- Authorization microservice for User management.
- Fleet microservice for vehicle management.
- Vehicle rooting microservice for Vehicle Routing Problems (VRP).
- Rooting machine microservice for vehicle rooting microservice.
- Geomessenger microservice for generating C-ITS messages.
- C-ITS broker microservice for broadcasting C-ITS messages.

The CTMaaS platform is comprised of two front-end modules: 1) the Web interface for the fleet manager and 2) the Mobile application for the driver. Concerning the back end, which is the database, this was developed also with microservices technologies. The microservices approach enables the use of various kinds of relational database management systems for different services. In the CTMaaS platform, two different Relational Database Management Systems (RDBMS) are used. The first database, Microsoft SQL Server, is embedded inside the Authentication microservice and keeps information related to user authentication, token lifespan, and general user profile information. The second RDBMS, called PostgreSQL, relates to the fleet microservice as it stores all data for trip planning, vehicle management, driver management, and vehicle positions. Considering the CTMaaS platform's need for geographic data management, PostGIS spatial extension was used on top of the already existing PostgreSQL RDBMS. PostGIS adds support for geographic objects allowing location queries to run in SQL.

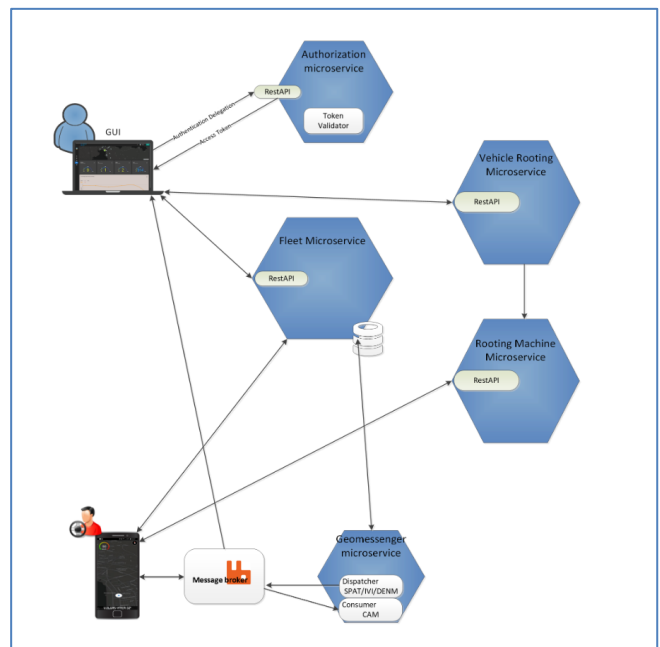

*Figure 3:* CTMaaS Architecture

The Fleet Microservice for vehicle management is the most essential service for the operation of the platform. Every procedure related to the vehicles' information, such as data entries, processing, and deletion is performed individually by this microservice. The assignment of vehicles to routes and destinations, the vehicles fleets' maintenance procedures, and the drivers' assignments are executed also by this microservice. The microservice includes various tables with information about the vehicles, their location, their condition, the trips they have to execute, the name of the driver, and the exact date of maintenance of the vehicle.

## *4. Conclusions and suggestions for future research*

This paper provides a thorough description of the CTMaaS platform and its goals, which include holistic traffic management, successful integration of C-ITS services, enhancement of the accessibility of an area through traffic management technologies, and improvement of everyday life. The C-ITS services integrated within the platform focus on fleet management needs. The mobile application developed within the project demonstrates real-time information and guidance to fleet managers and fleet drivers. A thorough description of the methodology used for the development of the CTMaaS platform and the steps followed are described. Every use case is analyzed to provide a detailed description of the scenarios the platform can perform. The design approach used for the development of the platform is that of Microservices.

The innovative approach of the CTMaaS project relies in combining the use of C-ITS services with the optimization of traffic management and control. What differentiates the CTMaaS platform from other similar platforms is the prediction of traffic conditions on the road and the direct communication with the Traffic Control Centers.

The literature review executed for the paper enabled the identification of various fleet management systems and services and highlights the next research directions. Emissions reduction and supporting environmental goals and policies are parameters that could be further analyzed and integrated into the CTMaaS platform. Regarding further research, another interesting topic is management services for individual vehicles or fleets of autonomous vehicles, using C-ITS services. The complexity of automated driving and the need to coexist with conventional vehicles on the road network can cause unfortunate events such as unsafe merges in the highway environment (Freddy Antony Mullakkal-Babu, 2020). To avoid such problems, research should be carried out on how a fleet of autonomous vehicles should operate, exploiting at the same time C-ITS technologies.

## *Acknowledgment*

The CTMaaS platform is the final product of an 18-month project called "Traffic Management as a Service by Implementing Cooperative Intelligent Transportation Systems - CTMaaS", which is funded by the Operational Program Central Macedonia 2014-2020.